\documentclass[12pt,a4paper]{article}
\pdfoutput=1
\usepackage[utf8]{inputenc}
\usepackage[T1]{fontenc}
\usepackage[top=2.5cm, bottom=2.5cm, left=2.0cm, right=2.0cm]{geometry}

\usepackage{amscd,amsfonts,amsmath,amssymb,amsthm,bbm,bm,latexsym,mathrsfs,mathtools,thmtools}
\usepackage[subnum]{cases}	
\allowdisplaybreaks			

\def\E {\mathbb{E}}
\def\V {\mathbb{V}}
\usepackage{scalerel}

\usepackage{hyperref} 		
\hypersetup{bookmarks=true, unicode=true, urlcolor=blue, linkcolor=blue, filecolor=blue, colorlinks=true, citecolor=black, final=true}

\usepackage[table]{xcolor}
\usepackage{subfigure,epsfig,graphics,graphicx,rotating} 
\usepackage{caption}
\usepackage{enumitem}
\usepackage{setspace}
\usepackage{longtable,float,array,multirow,booktabs}
\newcolumntype{P}[1]{>{\raggedright\arraybackslash}p{#1}}

\makeatother
\title{\vspace*{-50pt} Filtering the intensity of public concern from social media count data with jumps}

\author{
\hspace{-20pt}
Matteo Iacopini\textsuperscript{a}\thanks{e-mail: \href{mailto: m.iacopini@vu.nl}{m.iacopini@vu.nl}} \hspace{20pt} Carlo Santagiustina\textsuperscript{b}\thanks{e-mail: \href{mailto: carlo.santagiustina@unive.it}{carlo.santagiustina@unive.it}} \\ \\
\small \textsuperscript{a}Vrije Universiteit Amsterdam, The Netherlands\\
\small \textsuperscript{b}Ca' Foscari University of Venice, Italy
}

\date{\normalsize \today}

\begin{document}

\maketitle
\singlespacing

\begin{abstract}
Count time series obtained from online social media data, such as Twitter, have drawn increasing interest among academics and market analysts over the past decade.
Transforming Web activity records into counts yields time series with peculiar features, including the coexistence of smooth paths and sudden jumps, as well as cross-sectional and temporal dependence.
Using Twitter posts about country risks for the United Kingdom and the United States, this paper proposes an innovative state space model for multivariate count data with jumps.
We use the proposed model to assess the impact of public concerns in these countries on market systems. To do so, public concerns inferred from Twitter data are unpacked into country-specific persistent terms, risk social amplification events, and co-movements of the country series.
The identified components are then used to investigate the existence and magnitude of country-risk spillovers and social amplification effects on the volatility of financial markets.

\vspace*{10pt}
\end{abstract}

\section{Introduction}  \label{sec:introduction}

With globalisation, the need to understand the dynamics of country-risk perception and its effects on financial markets has become a relevant issue to investors, central bankers, and governments  for both portfolio diversification and debt issuance \cite{campbell2001have,hassan2003country,huber2019does}.
Major political events are known to be an important driver of market volatility \cite{bialkowski2008stock}; however, a measure of the underlying country-risk perception is still lacking.

The effect of new information on volatility in international markets has been studied extensively in the past few years. Previously, it was held that global risks are the only considerable risks in financial markets, but more recently, researchers have begun to examine country-risks and global risks to uncover local factors that cause stock market return volatility.
Moreover, options have been traded on the VIX index since 2006, thus allowing volatility to be considered an asset.

The identification of public concerns related to specific issues of interest (e.g., country risks) requires a data source that can be used to proxy the perception of the general public.
Unfortunately, existing sources of information on country-specific risks are either qualitative \cite{wef2017grr} or low frequency \cite{howell2011international} and often represent the perceptions of small panels of experts, thus ruling out the possibility of using them as proxies for the general public's concerns.

To address this gap, in this paper, we focus on count series generated from big textual data regarding country-specific risks via online social media. 
This type of data is characterised by sudden information avalanches and is affected by topic-specific trends, thus requiring adequate statistical tools to disentangle these components.
In this paper, we propose a new state space model for Web count data that allows country-specific risk series to be categorised into three separate components, each capturing a specific feature driving the intensity of the observables. 
In particular, (i) a smooth term represents the latent intensity of country-specific risk concerns, while (ii) a multiplicative jump component captures their social amplification, and (iii) a global component controls for common factors.
%
This method of decomposing country-risk perception offers new ways to evaluate the impact of social media phenomena on market and sociopolitical systems.
In this paper, we focus on the analysis of risk perception issues, but the proposed framework enables the use of count time series to analyse any Web phenomenon characterised by social amplification.


To overcome the limitations of the aforementioned data sources, we investigate user-generated Web data. 
Content generated by the Web community, such as online social media post-writing and sharing activities, have been widely used to track aggregate behaviour and infer the dynamics of the public's attention, perceptions, and concerns about specific topics of interest \cite{rogers2013debanalizing}.
Compared to other data sources (e.g., surveys and official statistics), big textual data from the Web allow for (i) larger sample sizes, (ii) higher velocity, and (iii) real-time collection \cite{varian2014big}.

Data from social media generally consist of a collection of character strings, which can be transformed into other data types for analysis with standard statistical methods \cite{einav2014data}.
For example, count data can be generated from texts through filtering conditions, which may consist of regular expressions (RegEx) or metadata restrictions (e.g., the author, geolocation, or date).
Count data from the Web are becoming increasingly valuable and widely used in several fields. For example, data from Google's search engine and Twitter have been used recently to improve forecasting accuracy in macroeconomics \cite{DAmuri2017Google_unemployment} and finance \cite{Ranco2015Twitter_stocks_PlosOne}.
The analysis of Web-based count data requires suitable and simple statistical tools designed for analysis and making predictions.

\begin{figure}[t!h]
\centering
\hspace*{-10pt}
\setlength{\abovecaptionskip}{4pt}
\begin{tabular}{cccc}
\begin{rotate}{90} \hspace{42pt} {\footnotesize US} \end{rotate} \hspace{-2.3ex} &
\includegraphics[trim= 25mm 2mm 22mm 5mm,clip,height= 3.5cm, width= 8.3cm]{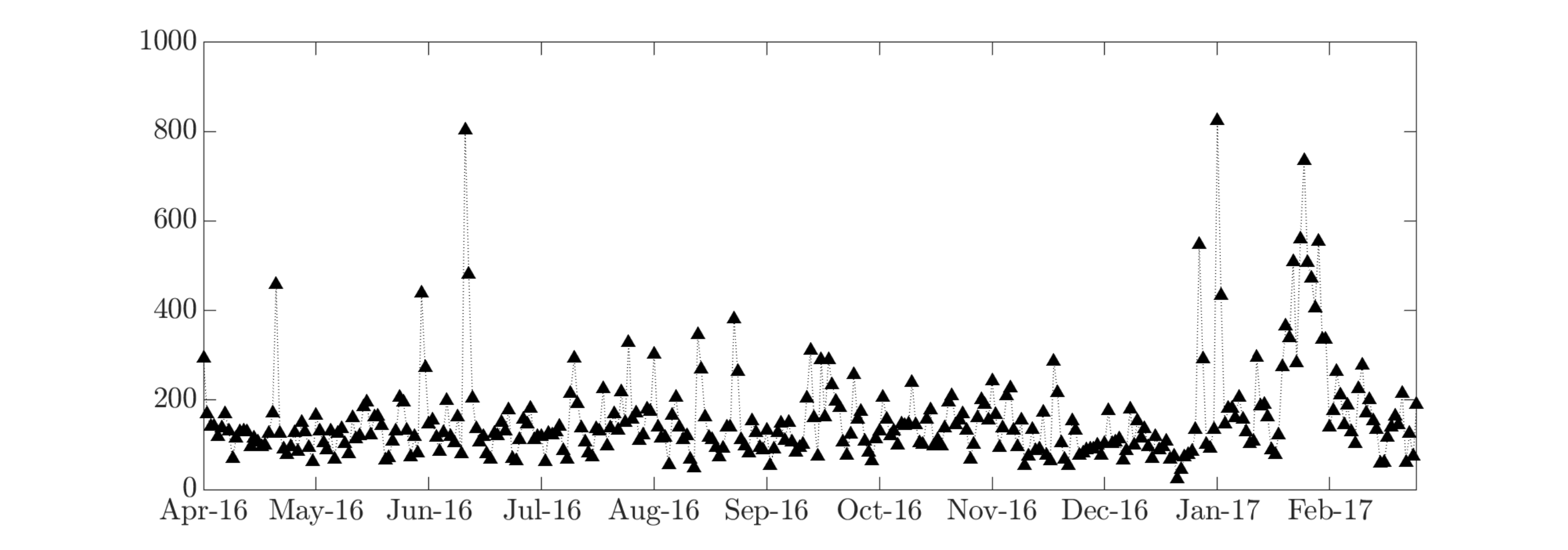} &
\begin{rotate}{90} \hspace{42pt} {\footnotesize UK} \end{rotate} \hspace{-2.3ex} &
\includegraphics[trim= 25mm 2mm 22mm 5mm,clip,height= 3.5cm, width= 8.3cm]{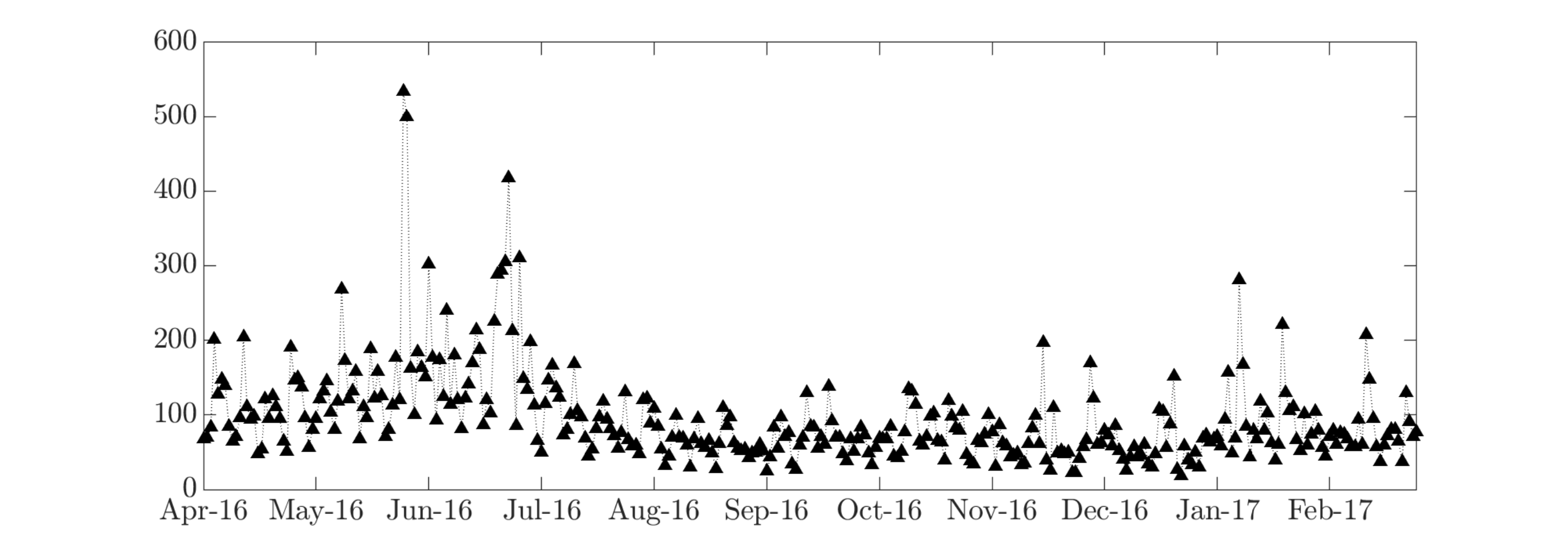} \\
\multicolumn{4}{c}{$\rho_{\scaleto{US}{4pt}}(1) = 0.50$, \qquad\;\; $\rho_{\scaleto{US}{4pt},\scaleto{UK}{4pt}} = 0.25$, \qquad\;\;  $\rho_{\scaleto{UK}{4pt}}(1) = 0.57$}
\end{tabular}
\caption{\footnotesize Top: daily time series of counts of Twitter posts from 1 April 2016 to 1 March 2017 for the US (left) and the UK (right). Bottom: auto-correlation of order 1 and instantaneous cross-correlation between US and UK.}
\label{fig:data_example}
\end{figure}

Figure \ref{fig:data_example} shows the daily number of Twitter posts published from April 2016 to March 2017 that contain the term ``\textit{risk}'' and refer to the United States (US) or to the United Kingdom (UK). The series share some peculiar features: (i) smooth evolution and (ii) several jump events.
%
The observed series are also characterised by positive auto- and cross-correlation, thus suggesting the presence of both temporal and cross-sectional dependence. 
Social media users' concerns regarding global-scale risks (e.g., trade wars, global financial crisis, and pandemics) may also impact country-specific activity, thus inducing the positive cross-correlation between the observed series in Fig. \ref{fig:data_example}.
To account for this fact, we enlarge our sample by including a third series regarding global risks, which are not imputable to a specific country, and use it as a common driver of US and UK observables.
State space (or parameter-driven) models provide a flexible and interpretable framework for the structural analysis of serially-dependent time series \cite{Durbin2012StateSpace_book}.
To account for the correlation structure of the data, we propose a new state space model in which the smooth temporal evolution of the observed variables is driven by a dynamic latent series representing social media users' concerns.

A parameter-driven Bayesian model for multivariate count data was recently proposed by \cite{Aktekin2018Bayes_SMC_count}. In their framework, the cross-sectional sum of past observations affects the dynamics of all latent series. 
Similarly, \cite{Zaman2014Bayes_predict_Twitter} proposed a method for predicting the time path of retweets. However, our goal is to decompose the public' latent concerns regarding country-specific risks.


An additional distinctive feature of the data in Fig.~\ref{fig:data_example} is the presence of country-specific and non-synchronised activity peaks. These sudden jumps cluster over time and have different degrees of persistence and scale.
Time series characterised by smooth trajectories with discontinuities that cluster over time are usually associated with exceptional events or structural breaks and call for the use of Markov-switching processes \cite{fruhwirth2006finite,chen2019markov}.

The sudden and transient nature of these jumping patterns allow them to be characterised as exceptional risk amplification events affecting the intensity of social media users' concerns.
Social amplification phenomena are known to occur on Web platforms, such as Twitter, especially for risk-related issues \cite{fellenor2018social}.
We define social amplification as the sudden intensification of online communications containing a specific risk-related theme (e.g., a country). See \cite{kasperson1988social} for a theoretical framework of the social amplification of risks.
The concerns of online communities can be amplified by salient events, online news coverage, and other sociocultural factors affecting risk perception \cite{renn1992social}.
Recently, \cite{strekalova2017beyond} found evidence of risk amplification in health-related risk debates occurring on Facebook. Similarly, \cite{wang2016spatial} analysed Twitter posts to identify the intensity of societal concerns regarding climate-change related risks.

Country risks are important in the explanation and prediction of market volatility indices \cite{ferreira2005have}. In particular, market operators are interested in identifying the specific contribution of each driver of public concerns in explaining the fluctuations of a volatility index to determine the country-risk spillover effects \cite{hoti2005modelling}. The measurement of country-risk perception may also empower government institutions with a new tool suitable for enacting their policy objectives, such as issuing of new government bonds.
Moreover, by disentangling the different components of country-related public concern and their relationship with volatility, it is possible to assess the relative contribution of rational and irrational behaviours in explaining volatility indices.
Motivated by these facts, in this paper, we are concerned with the identification and disentanglement of country risks, which are then used to explain the fluctuations of a market volatility index, the VIX.

The contributions of this paper are manifold.
First, we identify the timing, persistence, and scale of social amplification events.
Second, series-specific (country) effects are disentangled from common (global) contributions to intensity while controlling for exogenous factors that may contribute to explaining fluctuations in public concerns.
Finally, the extracted intensities are used to highlight country-specific contributions to fluctuations in the VIX index.

The reminder of the paper is organised as follows.
Section~\ref{sec:model} presents the new statistical model for count time series data. The inferential procedure and algorithms are then described in Section~\ref{sec:inference}.
Next, the proposed methodology is used in Section~\ref{sec:application} to study count data from Twitter posts that mention risk.
Finally, Section~\ref{sec:conclusions} summarises the main findings.

\section{A state space model for Web count data} \label{sec:model}


Despite widespread interest in analysing risk perception and social amplification in online social media, statistical tools capable of exploiting specificity in the aforementioned data are limited.
The peculiar features of count data obtained from social media, as discussed in the previous section, deserve special attention in the definition of a proper statistical model.
We note that counts extracted from Web data and our dataset share similar characteristics, thus making the proposed statistical framework applicable to a wide range of empirical studies.

Modelling time series of counts pose several challenges, such as discreteness of the observations, temporal and cross-sectional dependence and over-dispersion. Despite renewed interest over the past decade \cite{Weiss2018discrete_time_series_book}, these issues still need to be resolved.
%
Recent contributions include the dynamic Skellam model \cite{Koopman2017dynamic_Skellam} used for studying financial tick-by-tick data, \cite{yang2015state} study of health data with excess zeros, and the self-excited threshold Poisson model developed by \cite{Wang2014self_excited_Poisson}.
Within the Bayesian approach, \cite{park2011modelling} used a zero-inflated Poisson model to study counts of user generated content in an on-line community.
In this Section, we propose a novel state space model for multivariate count time series which allows for: serial and cross-sectional correlation, sudden signal amplification events which cluster over time, and smooth dynamics. Computationally, the proposed model scales linearly in the cross-sectional dimension.
We classify textual data from Twitter according to geographical markers (i.e., mention of the country in the text) to obtain a multivariate count time series $\{ y_{1,t},\ldots,y_{J,t},z_t \}_t$ consisting of country-specific series $y_{j,t}$ for each  $j=1,\ldots,J$, as well as a global series $z_t$ that stems from texts without a geographical reference to any of the selected countries.
See Section~\ref{sec:application} and the Supplement for further details on the construction of the dataset.

Starting from the count data illustrated in Fig.~\ref{fig:data_example}, we aim to extract and disentangle components of the underlying dynamic intensity driving the observed counts: the public's concerns for country risk.
As discussed in Section \ref{sec:introduction}, the existence of unobserved public concerns driving Twitter posting activity suggests the use of a state space framework for count time series \cite{Davis2016handbook_discrete_time_series}. This choice allows us to model data on their natural scale while preserving a direct interpretation of the components of the model.
We follow this approach and assume that the country-specific observables, $y_{j,t}$, are Poisson-distributed with a persistent, smooth latent intensity process, $x_{j,t}$, which represents country-risk concerns. This also allows for positive autocorrelation and instantaneous cross-correlation, as found in the data (see Section \ref{sec:introduction} and the preliminary analysis in Section~\ref{S_sec:application_preliminary} of the Supplement).



We account for persistent, smooth dynamics of positive-valued intensity $x_{j,t}$ by assuming a non-central Gamma distribution as the transition density of the process:
\begin{equation}
x_{j,t+1}|x_{j,t} \sim \text{NcGa}(\alpha_j, \beta_j x_{j,t},\delta_j).
\label{eq:latent_ARG}
\end{equation}
The non-central Gamma distribution is obtained as a Poisson mixture of Gamma. Hence if $x|z \sim \mathcal{G}a(a\!+\!z,c)$ with $z \!\sim\! \mathcal{P}oi(b)$, then $x \!\sim\! \textnormal{NcGa}(a,b,c)$, where $\mathcal{G}a(a,c)$ denotes a Gamma distribution with shape $a$ and scale $c$. It has density
\begin{equation*}
P(x) \!=\! \exp\Big(\!\!-\!\frac{x}{c}\Big) \! \sum_{k=0}^\infty \frac{x^{a+k-1}}{c^{a+k}\Gamma(a+k)} \frac{b^k \exp(-b)}{k!}, \quad x \in \mathbb{R}_+,\, a>0,\, b>0,\, c>0.
\end{equation*}
A non-central Gamma transition density defines an autoregressive Gamma process of order 1 \cite{Gourieroux06AR_GammaProcess}, or ARG(1).  
The parameter $\alpha_j$  in Eq. \eqref{eq:model_MS_NcGa_jumps_univariate} governs the magnitude of innovation in the latent state process, whereas $\beta_j,\delta_j$ account for persistence. \cite{Gourieroux06AR_GammaProcess} proved the stationarity of the ARG(1) process if $\beta_j \delta_j < 1$ and derived the conditional mean and variance as $\E[x_{j,t+1}|x_{j,t}] \!=\! \delta_j \alpha_j \!+\! \beta_j \delta_j x_{j,t}$ and $\V[x_{j,t+1}|x_{j,t}] \!=\! \delta_j^2 \alpha_j \!+\! 2 \beta_j \delta_j^2 x_{j,t}$.
The choice of the ARG process for the latent states is motivated by its flexibility and the fact that it facilitates the structural interpretation of the latent states (public concern in our empirical application).

As discussed in Section \ref{sec:introduction}, the sudden jumps magnifying the volume of observed counts in Fig. \ref{fig:data_example} are due to social amplification phenomena.
To avoid bias in the estimation of public concern, $x_{j,t}$, we introduce a positive multiplicative factor, $\xi_{j,t}$, thus obtaining the overall public concerns with amplification as
\begin{equation}
\tilde{x}_{j,t} = x_{j,t}(1+\xi_{j,t}) = x_{j,t} + x_{j,t}\xi_{j,t}.
\label{eq:latent_multiplicative_xi}
\end{equation}
The multiplicative factor allows us to disentangle the contribution of transient social amplification events, $x_{j,t}\xi_{j,t}$, from the persistent evolution of country-risk concerns, $x_{j,t}$.
Furthermore, because jumps in Twitter counts are persistent and cluster over time (see Section~\ref{sec:introduction}), we assume that country-specific social amplification is driven by an $L$-state hidden Markov chain $\lbrace s_{j,t} \rbrace_{t=1}^T$, such that $\xi_{j,t} \!=\! \xi_{j,s_{j,t}}$, where $\xi_{j,l}$ is a state-specific parameter.
The transition probabilities of each chain $j$ are assumed to be time-invariant and denoted by $\lambda_{j,l,k} \!=\! P(s_{j,t}\!=k | s_{j,t-1}\!=l)$, with transition matrix $\Lambda_j$.
Finally, to account for the effects of exogenous variables on latent intensity, we include country-specific covariates, $\mathbf{v}_{j,t}$, in the analysis.

The resulting univariate model with local Markov-switching jumps has a state space representation:
\begin{equation}
\begin{split}
y_{j,t}|x_{j,t},\boldsymbol{\xi}_j,s_{j,t} & \sim \mathcal{P}oi(x_{j,t}(1 +\xi_{j,s_{j,t}}) + \exp(\mathbf{v}_{j,t}'\boldsymbol{\phi}_j)) \\[2pt]
x_{j,t+1}|x_{j,t} & \sim \text{NcGa}(\alpha_j, \beta_j x_{j,t},\delta_j)
\end{split}
\label{eq:model_MS_NcGa_jumps_univariate}
\end{equation}
To assess the univariate model's \eqref{eq:model_MS_NcGa_jumps_univariate} ability to capture country-specific latent dynamics, we investigate the correlation structure of the observables counts and filtered latent intensities.
Table~\ref{tab:correlations} reports that the Spearman's rank correlation between the observed count time series is $0.257$, whereas it is $0.234$ between the latent intensities.
This result indicates that the univariate model is unable to correctly extract the latent intensity at country level since the correlation between the observables is entirely transferred to the latent intensities.
This is undesirable since we want latents to only capture the share of intensity specifically attributable to country-level features, removing the effect of common factors.
This calls for the use of a multivariate framework that accounts for positive correlation among the observables while allowing the impact of  co-movements and confounding factors to be disentangled from country-specific intensity.

\begin{table}[t!h]
\centering 
\footnotesize
\fbox{%
\begin{tabular}{*{3}{c} | *{2}{c} | *{2}{c}} 
\multicolumn{3}{c|}{\footnotesize observations} & \multicolumn{2}{c|}{\footnotesize latent multivariate} & \multicolumn{2}{c}{\footnotesize latent univariate} \\
$y_{\scaleto{US}{4pt},t},y_{\scaleto{UK}{4pt},t}$ & $y_{\scaleto{US}{4pt},t},z_t$ & $y_{\scaleto{UK}{4pt},t},z_t$ & $x_{\scaleto{US}{4pt},t},x_{\scaleto{UK}{4pt},t}$ & $\tilde{x}_{\scaleto{US}{4pt},t},\tilde{x}_{\scaleto{UK}{4pt},t}$ & $x_{\scaleto{US}{4pt},t},x_{\scaleto{UK}{4pt},t}$ & $\tilde{x}_{\scaleto{US}{4pt},t},\tilde{x}_{\scaleto{UK}{4pt},t}$ \\
\hline
0.257 & 0.447 & 0.387 & 0.066 & 0.071 & 0.234 & 0.249
\end{tabular} 
}%
\caption{\label{tab:correlations} Spearman's rank correlation among observed counts $(y_{j,t},z_t)$ and latent local intensities, without amplification ($x_{j,t}$) and with amplification ($\tilde{x}_{j,t}$).}
\end{table}

\begin{figure}[t!h]
\centering
\setlength{\abovecaptionskip}{7pt}
\includegraphics[trim= 0mm 0mm 0mm 0mm,clip,height= 5.0cm, width= 5.6cm]{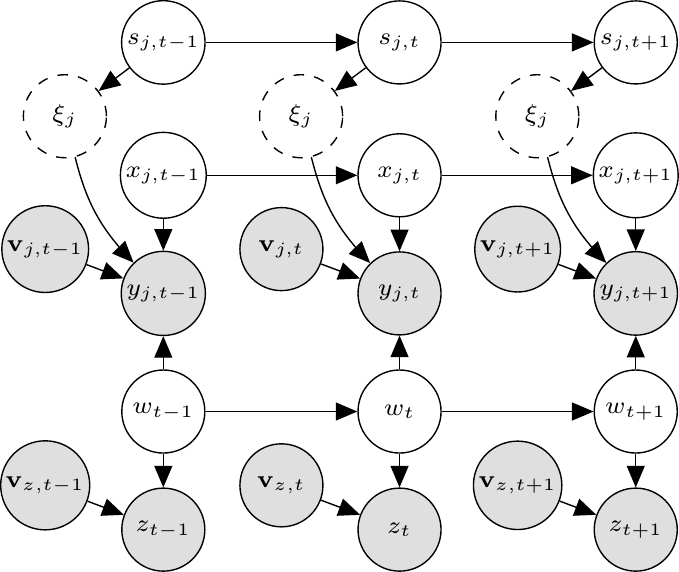}
\caption{{\protect\color{black}DAG of the model: observables (shaded gray circles), autoregressive components of the latent intensities and states (solid white circles), and amplification factors of latent intensity (dashed white nodes). Arrows' directions indicate the causal dependence relationships of the model.}}
\label{fig:flow_model_general}
\end{figure}

To account for the cross-sectional dependence structure among the $y_{j,t}$, we distinguish country-specific public concerns from global ones, and include a common latent factor, $w_t$, in country-level intensity.
The latent process $w_t$ encodes Twitter users' perceptions of global (rather than country-specific) risks; thus, it is assumed to drive the temporal evolution of the global series, $z_t$.
Note that including series $z_t$ helps in isolating the impact of common risk factors and seasonal effects from the country-specific concerns.
This specification leads to a multivariate model with local Markov-switching social amplification, whose conditional independence relationships are summarised by the DAG in Fig.~\ref{fig:flow_model_general}. The state space representation is
\begin{equation}
\begin{split}
z_{t}|w_t & \sim \mathcal{P}oi(w_t + \exp(\mathbf{v}_{z,t}'\boldsymbol{\phi}_z)) \\[2pt]
y_{j,t}|w_t,x_{j,t},\boldsymbol{\xi}_j,s_{j,t} & \sim \mathcal{P}oi(w_t + x_{j,t}(1 +\xi_{j,s_{j,t}}) + \exp(\mathbf{v}_{j,t}'\boldsymbol{\phi}_j)) \\[2pt]
x_{j,t+1}|x_{j,t} & \sim \text{NcGa}(\alpha_j, \beta_j x_{j,t},\delta_j) \\[2pt]
w_{t+1}|w_{t}     & \sim \text{NcGa}(\alpha_w, \beta_w w_{t},\delta_w)
\end{split}
\label{eq:model_MS_NcGa_jumps}
\end{equation}
We refer to the components of country-level intensity as follows: `local' $x_{j,t}$, `amplification' $x_{j,t} \xi_{j,s_{j,t}}$, `global' $w_t$, and `covariates', $\exp(\mathbf{v}_{j,t}'\boldsymbol{\phi}_j)$.

{\color{black}%
The contribution of the proposed model for multivariate count time series is manifold.
First, it accounts for series-specific and global latent intensities driving the observed counts, allowing to disentangle the impact of common factors from idiosyncratic latent dynamics.
Jumps have been mainly studied in the context of real-valued time series \cite{andersen2007roughing} using an additive specification. Instead, we allow for jumps in count time series through a series-specific multiplicative term which amplifies the effect of series-specific intensities. Despite their limited use in the literature \cite{caporin2017chasing}, series-specific multiplicative jumps allow for a more flexible modelling as compared to the widespread additive specification.
Finally, we do not make the temporal independence assumption for the jump terms and model them using a hidden Markov chain that allows to infer different states of excitation within the time series of counts.

Besides, the functional form of model \eqref{eq:model_MS_NcGa_jumps} permits a direct structural interpretation of the latent variables and parameters as series-specific or global intensities, and amplification.
This is granted by the assumption of an ARG process for the latent intensities. Differently, the commonly used lognormal specification \cite{wang2018modelling,heinen2007multivariate}, despite allowing the inclusion of similar terms in the intensity, would make the interpretation more difficult due to the nonlinear exponential link function.
}%

\cite{Jorgensen1999StateSpace_count} presented a state space for multivariate Poisson data where the observed counts are driven by a common Markov process. However, our framework is more general in two respects. First, we assume a different state transition that allows for both stationary and nonstationary processes. Second, our inferential procedure also holds for different observation densities.
Recently, \cite{chen2019markov} proposed a Markov-switching Poisson integer-valued GARCH model to account for consecutive zeros and time-varying volatility in count data.
Since their model is univariate and observation-driven, which means that the latent intensities are perfectly predictable, it is unable to account for features of the data highlighted in Section \ref{sec:introduction}.
Alternative approaches for multivariate count time series have been developed by \cite{wang2018modelling}, who modelled dependence by assuming a factor structure for latent intensities, and \cite{heinen2007multivariate}, who exploited copulas to impose dependence among the observables while assuming a VARMA process for the dynamics of log-intensities.

From a forecasting perspective, \cite{berry2020bayesian} used the decouple/recouple strategy to define a dynamic model for multivariate counts that is computationally scalable in the number of series.
Conversely, our interest lies in identifying and disentangling the structural components of latent public concerns. This motivates the parametric assumptions underlying our model \eqref{eq:model_MS_NcGa_jumps}, which prevent the direct applicability of their decouple/recouple strategy.


{\color{black}
Given the structural form of model \eqref{eq:model_MS_NcGa_jumps}, the likelihood function is a high-dimensional integral with no closed-form solution. Hence, we apply a data augmentation approach and introduce the latent state variables $s_{j,t},x_{j,t},w_t$ in the set of observations, thus obtaining the complete-data likelihood function
\begin{align}
\notag
& L(\mathbf{Y},\mathbf{Z},\mathbf{V},\mathbf{S},\mathbf{W},\mathbf{X}|\boldsymbol{\theta}) = \prod_{t=1}^T \text{NcGa}(\alpha_w, \beta_w w_{t-1},\delta_w)  \prod_{t=1}^T \prod_{j=1}^J \text{NcGa}(\alpha_j, \beta_j x_{j,t-1},\delta_j) \\ \notag
 & \cdot \prod_{l=1}^L \prod_{j=1}^J \prod_{t \in \mathcal{T}_{j,l}} \! \frac{(w_t \!+\! x_{j,t}(1 \!+\! \xi_{j,l}) \!+\! \exp(\mathbf{v}_{j,t}'\boldsymbol{\phi}_j))^{y_{j,t}}}{y_{j,t}!} \exp(-(w_t \!+\! x_{j,t}(1 \!+\! \xi_{j,l}) \!+\! \exp(\mathbf{v}_{j,t}'\boldsymbol{\phi}_j))) \\
 & \cdot \prod_{t=1}^T \frac{(w_t + \exp(\mathbf{v}_{z,t}'\boldsymbol{\phi}_z))^{z_t}}{z_t!} \exp(-(w_t + \exp(\mathbf{v}_{z,t}'\boldsymbol{\phi}_z))) \prod_{j=1}^J \prod_{l=1}^L \prod_{k=1}^L \lambda_{j,l,k}^{N_{lk}(\mathbf{S}_j)} p(\mathbf{s}_{j,0}|\Lambda_j),
\label{eq:complete_data_likelihood}
\end{align}
where $\mathcal{T}_{j,l} = \{ t: s_{j,t}=l \}$ and $N_{lk}(\mathbf{S}_j) = \#\{ s_{j,t-1} = l, s_{j,t} = k \}$ are the number of transitions from state $l$ to state $k$ of the $j$-th chain $\mathbf{S}_j$.
Motivated by our interest in identifying of social amplification phenomena in data, we assume $L=2$ and impose $\xi_{j,1} = 0$ for every $j$. This leads to identifying regime 1 as having no jumps and regime 2 as the social amplification regime.
However, the model specification is general and allows for any number of regimes $L > 1$ that can be identified using the constraint $\xi_{j,1} < \ldots < \xi_{j,L}$.
}%

\section{Bayesian inference}  \label{sec:inference}
\subsection{Prior specification}

In this paper, we adopt a Bayesian approach that provides a simple way to introduce regularisation via the specification of appropriate prior distributions and permits a flexible modelling of jumps through hierarchical priors.
The parameters $\alpha_j$ and $\delta_j$ (similarly, $\alpha_w$ and $\delta_w$) of the non-central Gamma distribution in \eqref{eq:model_MS_NcGa_jumps} are not separately identifiable, thus requiring the introduction of constraints for their estimation.
To solve the identification issue, we introduce a soft constraint by specifying a truncated prior distribution for $\delta_j,\delta_w$, which bounds them away from zero, and then test the robustness of the results for various truncation levels.
In all the analyses performed, we find that the parameters $\delta_j,\delta_w$ are never stuck at the lower bound, thus providing evidence that the proposed constraint is non-binding. 
See the Supplement for further details regarding the proposed truncation scheme and values of the hyper-parameters.

\begin{figure}[t!h]
\setlength{\abovecaptionskip}{7pt}
\centering
\includegraphics[trim= 0mm 0mm 0mm 0mm,clip,height= 5.8cm, width= 9.3cm]{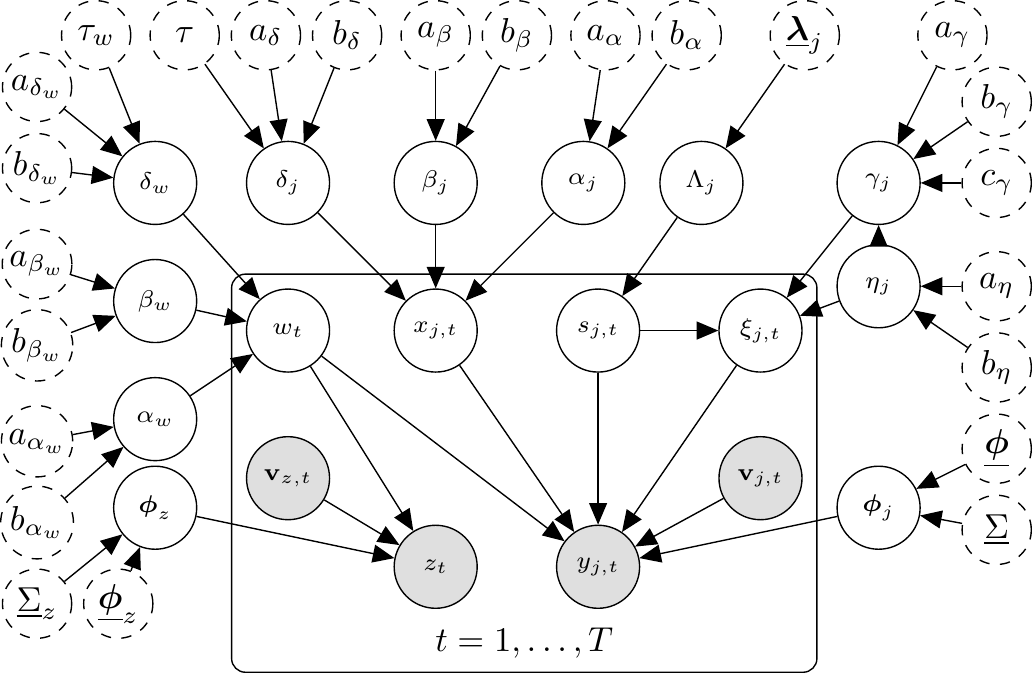}
\caption{DAG of model \protect\eqref{eq:model_MS_NcGa_jumps}. It exhibits the hierarchical structure of priors and related hyper-parameters: observables (shaded, gray circles), latent parameters (solid, white circles), and fixed hyper-parameters (dashed, white circles). Arrows' directions indicate the causal dependence relationships of the model.}
\label{fig:flow_model_prior}
\end{figure}

We assume a flexible hierarchical prior for the country-specific social amplification factor, $\xi_{j,2}$. The overall prior structure for each $j=1,\ldots,J$ is 
\begin{alignat}{3}
\notag
& \eta_{j} \sim \mathcal{G}a(a_\eta,b_\eta) \qquad && \gamma_{j} | \eta_{j} \propto \dfrac{a_\gamma^{\gamma_{j}-1} \eta_{j}^{\gamma_{j} c_\gamma}}{\Gamma(\gamma_{j})^{b_\gamma}} \qquad && \xi_{j,2} | \gamma_{j}, \eta_{j} \sim \mathcal{G}a(\gamma_{j}, 1/\eta_{j}), \\[0pt] \label{eq:prior_model_general}
& {\color{black} \boldsymbol{\phi}_j \sim \mathcal{N}(\underline{\boldsymbol{\phi}}, \underline{\Sigma})} \qquad && {\color{black} \boldsymbol{\phi}_z \sim \mathcal{N}(\underline{\boldsymbol{\phi}}_z, \underline{\Sigma}_z)} && {\color{black} \boldsymbol{\lambda}_{j,l} \sim \mathcal{D}ir(\underline{\boldsymbol{\lambda}}_j)} \\[0pt] \notag
& \alpha_j \sim \mathcal{G}a(a_\alpha,b_\alpha) \qquad && \beta_j \sim \mathcal{G}a(a_\beta,b_\beta) \qquad && \delta_j \sim T\mathcal{G}a(a_\delta,b_\delta;S_\tau), \\[0pt] \notag
& \alpha_w \sim \mathcal{G}a(a_{\alpha_w},b_{\alpha_w}) \qquad && \beta_w \sim \mathcal{G}a(a_{\beta_w},b_{\beta_w}) \qquad && \delta_w \sim T\mathcal{G}a(a_{\delta_w},b_{\delta_w};S_{\tau_w}).
\end{alignat}
The notation $T\mathcal{G}a(a,b;\tau)$ stands for a Gamma distribution truncated on the interval $S_\tau = (0,\tau) \subset \mathbb{R}_+$, parametrised by $\tau$.
{\color{black} Finally, the unnormalised prior distribution for $\gamma_j$ is conjugated for the shape parameter of a Gamma distribution \cite{llera2016bayesian} with positive real  hyper-parameters $a_\gamma, b_\gamma, c_\gamma$. The conjugacy property allows for efficient posterior sampling via the inverse transform method without the need for a tuning parameter.
}%
Fig.~\ref{fig:flow_model_prior} reports the directed acyclic graph (DAG) of the model and prior structure.

\subsection{Posterior approximation}


The joint posterior distribution of parameters and latent variables is
\begin{equation*}
P(\boldsymbol{\theta},\mathbf{S},\mathbf{W},\mathbf{X} | \mathbf{Y},\mathbf{Z},\mathbf{V})  = P(\boldsymbol{\theta}) L(\mathbf{Y},\mathbf{Z},\mathbf{V},\mathbf{S},\mathbf{W},\mathbf{X}|\boldsymbol{\theta}).
\end{equation*}
Since this distribution is not tractable, we develop a MCMC algorithm to generate random draws from the posterior distribution and approximate all posterior quantities of interest. We refer to Appendix~\ref{sec:apdx_posterior_jumps_general} and the Supplement for the detailed derivation of posterior distributions.

Estimating the trajectory of latent intensity is known as the filtering problem. For special classes of state space models, such as linear Gaussian state space models, the filtering distribution is known in closed form, and efficient algorithms are available (e.g., the Kalman and Hamilton filters). In more general frameworks such as eq. \eqref{eq:model_MS_NcGa_jumps}, when no closed form expressions are available, the filtering problem may be solved by resorting to simulation-based methods, such as particle filters.
We follow this strategy and rely on the selection/mutation (SM) algorithm \cite[e.g., see][]{Cappe2005HMM} to obtain filtered estimates of the paths of latent states $w_t$ and $x_{j,t}$.
The SM algorithm addresses the degeneracy issue by selecting, at each point in time, a set of particles based on the associated weights and then simulating an independent extension for each selected trajectory.
%
The static parameters governing the dynamics and amplification factors are sampled using an adaptive Metropolis-Hastings (aRWMH) step \cite{Atchade2005adaptiveMH}. Finally, hyper-parameters are directly sampled from their corresponding posterior full conditional distributions.

We use the convention $\mathbf{X}_j \!=\! \{ x_{j,t} \!:\! t \!=\! 1,\ldots,T \}$ and $\mathbf{X} \!=\! \{ \mathbf{X}_{j} \!:\! j \!=\! 1,\ldots,J \}$ and similarly for the other variables and parameters. The MCMC algorithm is articulated in two main blocks, drawing the path of global intensity ($w_t$) and country-specific components ($x_{j,t}$, $s_{j,t}$, $\xi_{j,s_{j,t}}$), as follows:
\begin{enumerate}[label=(\arabic*)]
\item Sample the global latent intensity and the static parameters:
\begin{enumerate}[label=(\arabic{enumi}\alph*)]
\item Sample the latents $\mathbf{W}$ conditionally on  $(\mathbf{Y},\mathbf{Z},\mathbf{V},\alpha_w, \beta_w, \delta_w, \mathbf{X}, \boldsymbol{\xi}, \mathbf{S})$ using a particle filter with the SM algorithm;
\item Sample the static parameters of the ARG process for $w_t$:
\begin{itemize}
\item sample $\alpha_w$ from $P(\alpha_w | \mathbf{W}, \beta_w, \delta_w)$ via aRWMH;
\item sample $\beta_w$ from $P(\beta_w | \mathbf{W}, \alpha_w, \delta_w)$ via aRWMH;
\item sample $\delta_w$ from $P(\delta_w | \mathbf{W}, \alpha_w, \beta_w)$ via aRWMH;
\end{itemize}
\item Sample the coefficients $\boldsymbol{\phi}_z$ from $P(\boldsymbol{\phi}_z|\mathbf{Z},\mathbf{V}_z,\mathbf{W})$ via aRWMH;
\end{enumerate}

\item Independently for each $j=1,\ldots,J$, sample the country-specific latent intensity, hidden Markov chain, and static parameters:
\begin{enumerate}[label=(\arabic{enumi}\alph*)] 
\item Sample the latents $\mathbf{X}_j$ conditionally on $(\mathbf{Y}_j,\mathbf{V}_j,\alpha_j,\beta_j, \delta_j, \boldsymbol{\xi}_j, \mathbf{W}, \mathbf{S}_j)$ using a particle filter with the SM algorithm;
\item Sample the hidden chain $\mathbf{S}_j$ conditionally on $(\mathbf{Y}_j,\mathbf{V}_j,\mathbf{W}, \mathbf{X}_j, \boldsymbol{\xi}_j, \Lambda_j)$ using the FFBS algorithm;
\item Sample the rows of transition matrix $\Lambda_j$ from $P(\boldsymbol{\lambda}_{j,l}|\mathbf{S}_j)$ for $l=1,2$;
\item Sample, in block, the parameters associated with the jump terms:
\begin{itemize}
\item sample $\eta_{j}$ from $P(\eta_{j} | \boldsymbol{\xi}_{j}, \gamma_{j})$;
\item sample $\gamma_{j}$ from $P(\gamma_{j} | \boldsymbol{\xi}_{j}, \eta_{j})$ via the inverse transform method;
\item sample the amplification $\xi_{j,2}$ from $P(\xi_{j,2} | \mathbf{Y}_j, \mathbf{X}_j, \mathbf{W}, \gamma_{j}, \eta_{j}, \mathbf{S}_{j})$ via aRWMH;
\end{itemize}
\item Sample the static parameters of the ARG process for $x_{j,t}$:
\begin{itemize}
\item sample $\alpha_j$ from $P(\alpha_j | \mathbf{X}_j, \beta_j, \delta_j)$ via aRWMH;
\item sample $\beta_j$  from $P(\beta_j | \mathbf{X}_j, \alpha_j, \delta_j)$ via aRWMH;
\item sample $\delta_j$ from $P(\delta_j | \mathbf{X}_j, \alpha_j, \beta_j)$ via aRWMH;
\end{itemize}
\item Sample the coefficients $\boldsymbol{\phi}_j$ from $P(\boldsymbol{\phi}_j|\mathbf{Y}_j,\mathbf{V}_j,\mathbf{X}_j,\mathbf{S}_j,\boldsymbol{\xi}_j)$ via aRWMH.
\end{enumerate}
\end{enumerate}
We tested the sampler's performance in simulated experiments and compared it to a log-Normal specification for latent intensities $w_t,x_{j,t}$.
The results suggest that our framework is better suited for minimising the risk of miss-classifying ordinary fluctuations of the observables as social amplification events, which is one of the objectives in our application.
See the Supplement for further details.
Note that model \eqref{eq:model_MS_NcGa_jumps} and the proposed sampler can be easily modified to analyse binary or integer-valued time series since the observational densities in Eq. \eqref{eq:model_MS_NcGa_jumps} only affect the distribution to be approximated by the particle filter.

\section{Empirical application} \label{sec:application}
In this application we model the public concern among Twitter users regarding country-related risks in the UK and US.
According to \cite{chung2011social}, `{\textit{easy access to and efficient sharing of risk information [...] could accelerate the intensity as well as the speed of social attention to risk issues}}' .  Accordingly, we apply model \eqref{eq:model_MS_NcGa_jumps} to count data generated from Twitter posts mentioning risk to extract the intensity of country-risk concerns (e.g., political, macroeconomic, financial, or environmental) and  identify risk amplification events.

The period of investigation, which ranges from the 1 of April 2016 to the 1 of March 2017, is characterised by the United Kingdom's European Union membership referendum (23 of June 2016), the United States presidential elections (8 of November 2016), and their aftermath, including the Brexit negotiations and the Trump administration taking office. Referendums and presidential elections are major political events, so they receive broad media coverage and are characterised by intense speculation on their outcome and consequences.
Since these events can potentially generate public concern regarding their implications at the national and international level, this dataset provides an ideal setting for disentangling the different components of country risks and evaluating their impact on the volatility of financial markets.
%

\subsection{Data collection}
Tweets containing the word \textit{risk} were collected by programmatically querying the Search API of Twitter \cite{Makice2009Twitter}.
Further details about the the data strategy and query parameters are included in Section \ref{S_sec:data} of the Supplement. 
For each day of the period under investigation, we count the number of Twitter posts (considering both tweets and retweets) matching specific country-dictionary conditions (see RegEx  Tables~\ref{tab:dictionary_us}-\ref{tab:dictionary_uk} in the Supplement), which are used to identify tweets referring to country-specific risks affecting either the US or UK.  
By doing so, we obtain the count time series $y_{US,t}$ and $y_{UK,t}$.
Note that filtering risk data by country dictionaries without focusing on specific subtopics offers a broad lens to view all the (potentially time-varying) salient dimensions of country-risk perception.
Conversely, some of these dimensions may be missed when using topical dictionaries since they cannot provide, \textit{a-priori}, an exhaustive list of all the drivers of country risks.

Moreover, while country dictionary conditions (e.g., country names and acronyms) are exhaustive, unambiguous and stable over time, topic dictionary conditions with the same qualities are much more difficult to construct.
Similarly, we create the global series $z_t$ by counting tweets about risk(s) that do not refer directly to the US or UK  (i.e., tweets that do not match either country's dictionary conditions).
Tweets counted the global series may refer to risks related to policies (e.g., Trump's foreign policy),  events (e.g., the Brexit) and international factors (e.g., protectionism and international trade wars) which may affect both countries.

To control for the possible effect of breaking news and scheduled events that are expected to affect the volatility of financial markets, we also include (i) the share of newspaper articles containing the word \textit{risk}, at both the country and global level, and (ii) the country-specific number of expected high-volatility events as covariates. See Section \ref{S_sec:data} of the Supplement for further details.


\subsection{Results}
Our data consist of $T=334$ daily observations for $J=2$ countries (the US and UK), and a global series.
The count series for the UK and US are rescaled by a factor of $0.1$ and the global series by a factor of $0.001$.
We run the Gibbs sampler for $20,000$ iterations after discarding the first $4,000$ as burn-in.

\begin{figure}[h!t]
\centering
\hspace*{-8pt}
\setlength{\abovecaptionskip}{3pt}
\begin{tabular}{ccc}
\begin{rotate}{90} \hspace*{38pt} {\footnotesize US} \end{rotate} \hspace*{-3.0ex} & 
\includegraphics[trim= 50mm 0mm 38mm 0mm,clip,height= 3.4cm, width= 14.5cm]{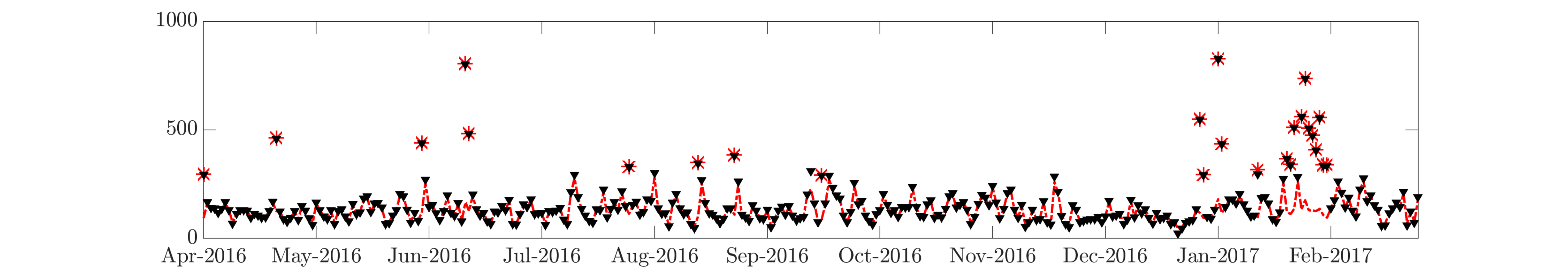} & \hspace*{-17pt}
\includegraphics[trim= 0mm 0mm 5mm 0mm,clip,height= 3.4cm, width= 2.9cm]{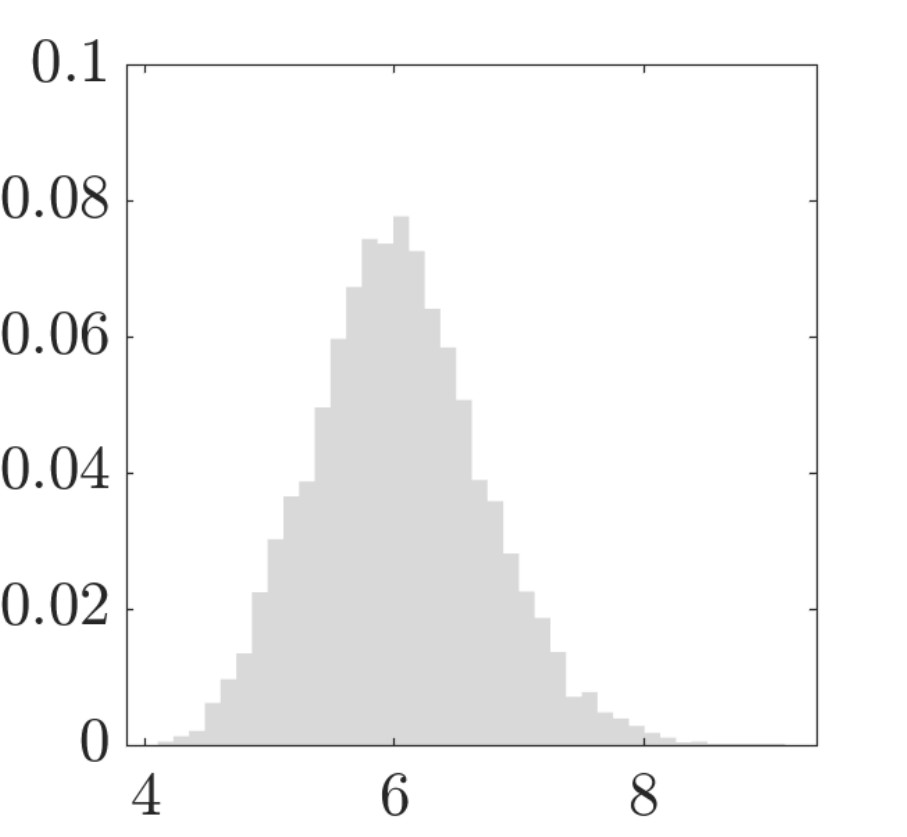} \\
\begin{rotate}{90} \hspace*{38pt} {\footnotesize UK} \end{rotate} \hspace*{-3.0ex} & 
\includegraphics[trim= 50mm 0mm 38mm 0mm,clip,height= 3.4cm, width= 14.5cm]{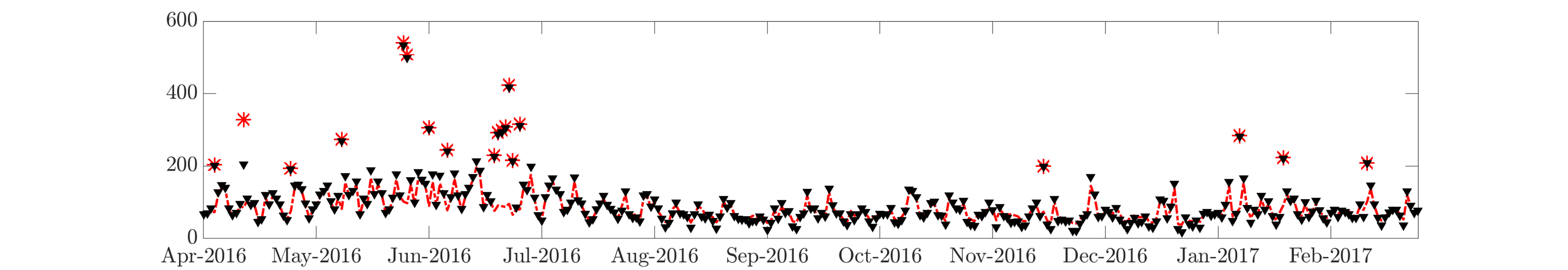} & \hspace*{-17pt}
\includegraphics[trim= 0mm 0mm 5mm 0mm,clip,height= 3.4cm, width= 2.9cm]{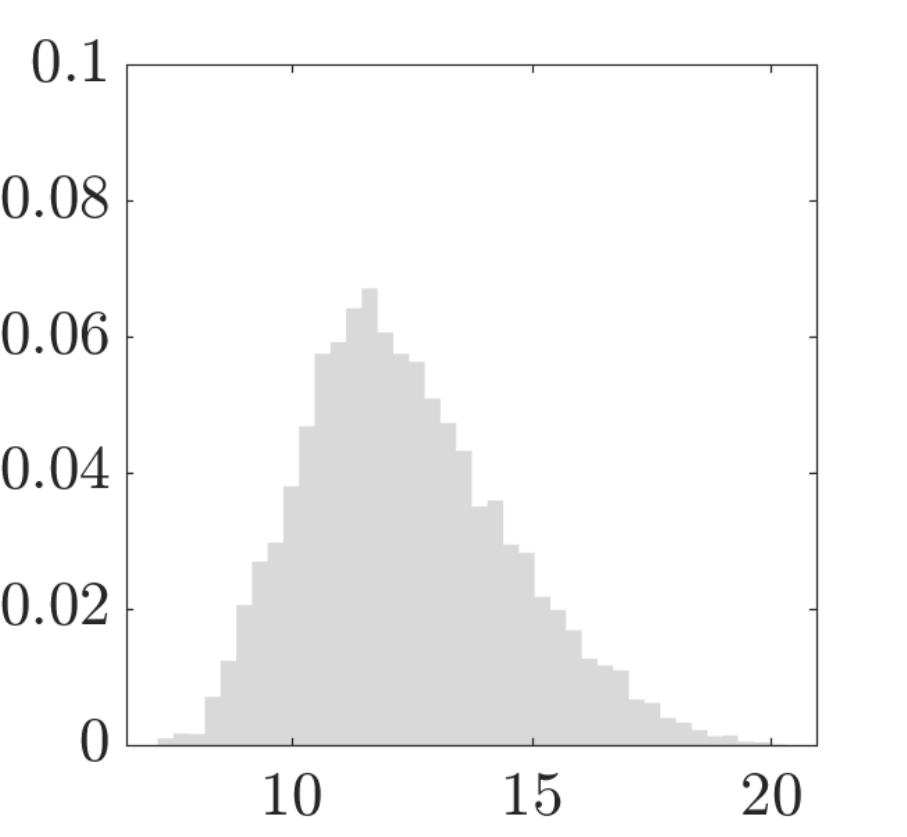}
\end{tabular}
\caption{{\protect\color{black}\textit{Left:} observed data (black triangles) and the posterior mean of filtered states (dashed red lines) and jumps (red stars).
\textit{Right:} posterior distribution of the social amplification factor.}}
\label{fig:app_risk_a0d08_filter}
\end{figure}

Figure~\ref{fig:app_risk_a0d08_filter} shows the original data and the estimated latent intensity, distinguishing the persistent country-specific part (red line) from the estimated social amplification phenomena (red stars).
The sampler identifies jumps that correspond to days characterised by the highest variation rates in the observations.
In the first four months, which include the Brexit referendum, the path of the persistent component for the UK experienced a tumultuous period with higher average values compared to the following months. 
For the US, latent intensity appears rather stable, with some turmoil from mid-July to the US elections in November, as Donald Trump's polling against Hillary Clinton rose steadility \cite{bovet2018validation}, as well as when he took office in late January 2017.

\begin{figure}[h!t]
\centering
\setlength{\abovecaptionskip}{3pt}
\begin{tabular}{ccc}
\begin{rotate}{90} \hspace*{39pt} {\footnotesize US} \end{rotate} \hspace*{-3.0ex} & 
\includegraphics[trim= 50mm 0mm 32mm 0mm,clip,height= 3.4cm, width= 14.0cm]{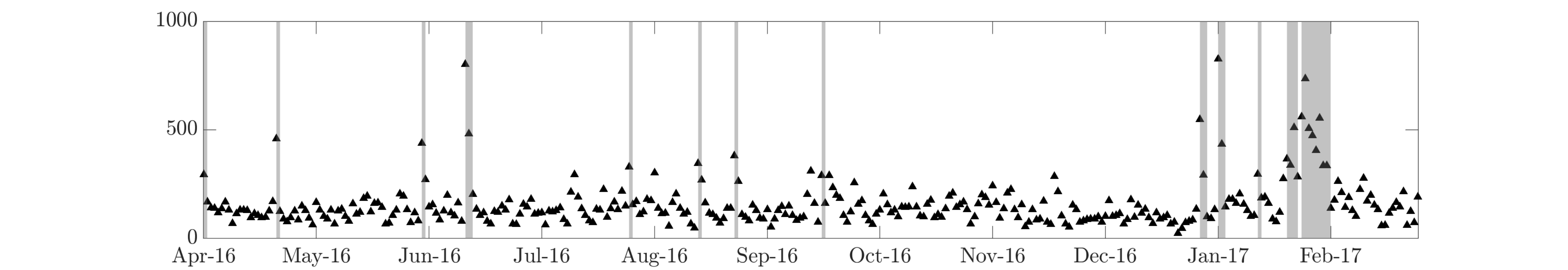} & \hspace*{-17pt}
\includegraphics[trim= 0mm 0mm 5mm 0mm,clip,height= 3.4cm, width= 2.9cm]{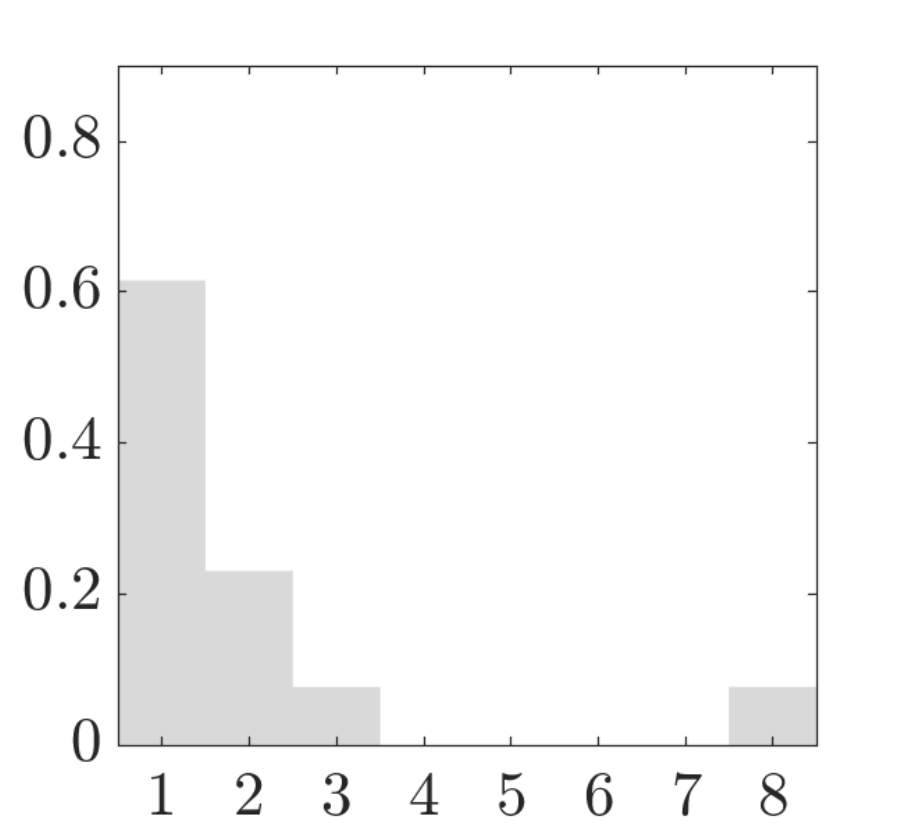} \\
\begin{rotate}{90} \hspace*{39pt} {\footnotesize UK} \end{rotate} \hspace*{-3.0ex} & 
\includegraphics[trim= 50mm 0mm 32mm 0mm,clip,height= 3.4cm, width= 14.0cm]{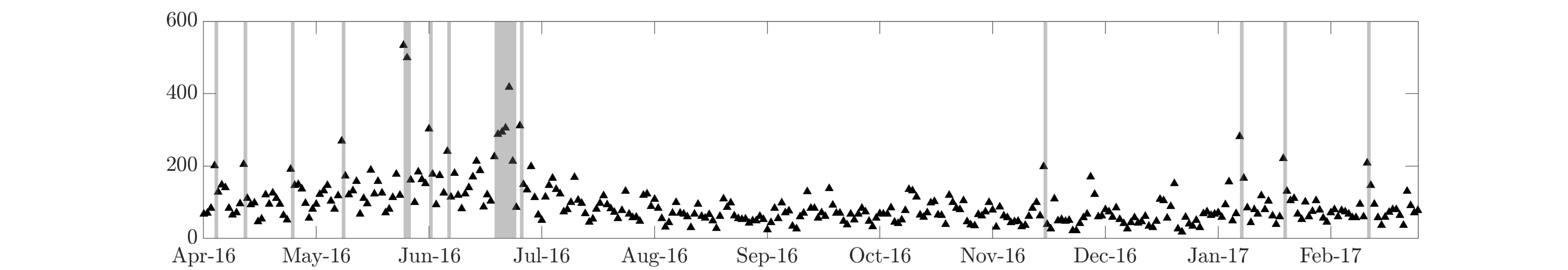} & \hspace*{-17pt}
\includegraphics[trim= 0mm 0mm 5mm 0mm,clip,height= 3.4cm, width= 2.9cm]{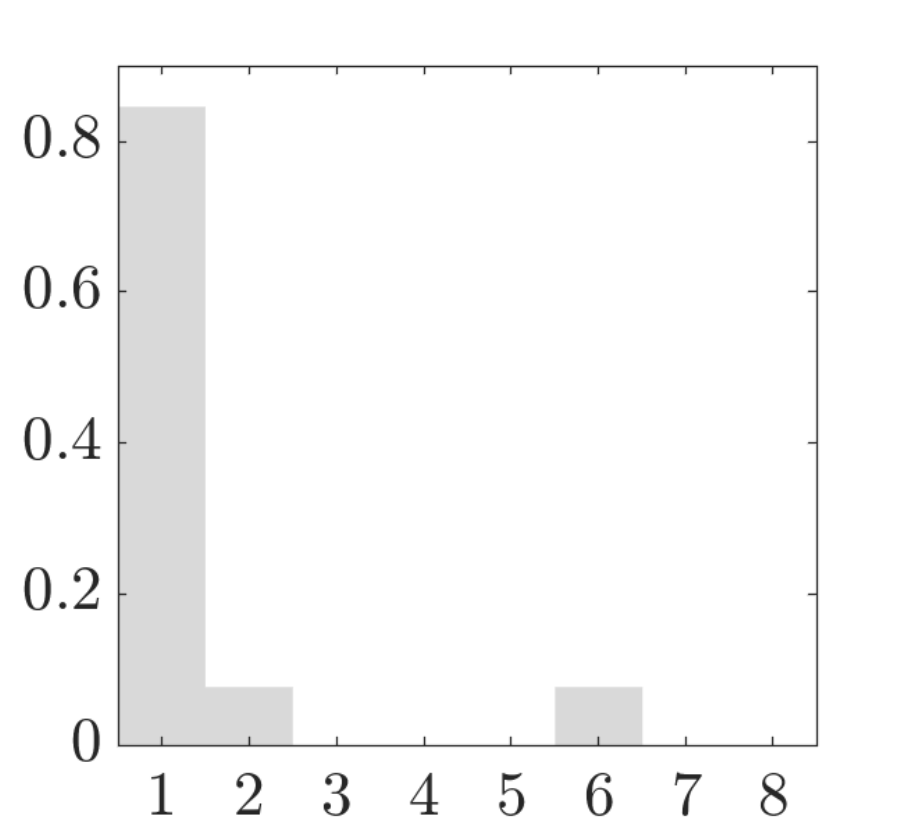}
\end{tabular}
\caption{{\protect\color{black} \textit{Left:} observation (black triangles) and estimated hidden chain $s_{j,t}$ or US and UK: regime 2 (grey bars) and regime 1 (white bars).
\textit{Right:} distribution of the duration of the social amplification regime (regime 2).}}
\label{fig:app_risk_a0d08_states}
\end{figure}

The days in which social amplification phenomena occur are reported as grey shades in Fig.~\ref{fig:app_risk_a0d08_states}, along with the distribution of their duration.
The different country-level results corroborate the interpretation of social amplification as a local and unsynchronised phenomenon. In particular, most UK events occur before and right after the Brexit referendum, whereas social amplification in the US is more concentrated when Donald Trump took office.
In particular, we find that during the first three months of the sample, corresponding to the period immediately prior to the EU referendum, which was characterised by the the Brexit victory and resignation of PM David Cameron, social amplification of risk with respect to the UK was frequent.

However, we find that events of social amplification are relatively more frequent for the US (i) at the beginning of June 2016, (ii) in August 2016, and (iii) from late December 2016 to February 2017. As aforementioned, interval (ii) was characterised by temporary reductions in the gap between Clinton and Trump in the polls, whereas the interval (iii) follows Donald Trump taking office as the President of the United States, and corresponds to the commencement of Trump's ``America First'' domestic and foreign policy.

Moreover, from the right-hand column of Figs.~\ref{fig:app_risk_a0d08_filter}-\ref{fig:app_risk_a0d08_states}, we find that the average duration of amplification events in the UK is lower compared to the US, but their amplitude is higher in relative terms, capturing the fact that social amplification in the UK is more exceptional and intense compared to the US. This provides evidence in favour of the interpretation of social amplification as a transient phenomenon, which rapidly dissipates without having any persistent effect on the public's perception of risk. 


\begin{figure}[h!t]
\centering
\setlength{\abovecaptionskip}{0pt}
\begin{tabular}{cc}
\begin{rotate}{90} \hspace*{39pt} {\footnotesize US} \end{rotate} \hspace*{-13pt} &
\includegraphics[trim= 50mm 0mm 32mm 0mm,clip,height= 3.5cm, width= 17.0cm]{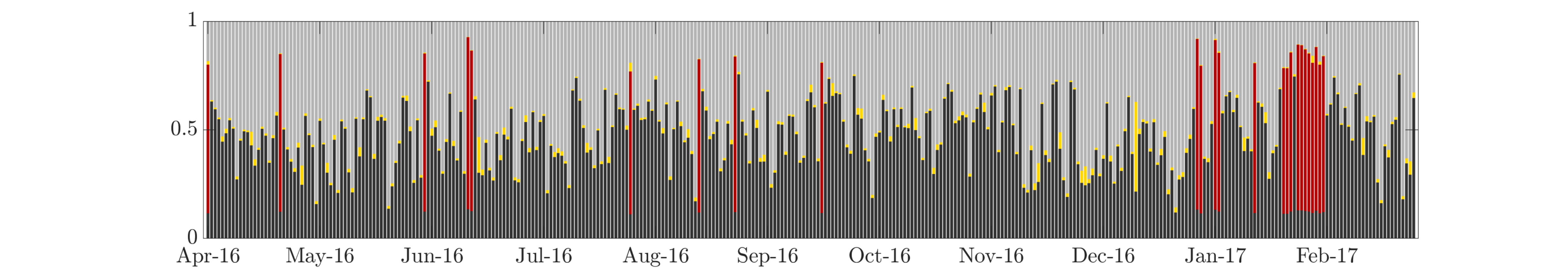} \\
\begin{rotate}{90} \hspace*{39pt} {\footnotesize UK} \end{rotate} \hspace*{-13pt} &
\includegraphics[trim= 50mm 0mm 32mm 0mm,clip,height= 3.5cm, width= 17.0cm]{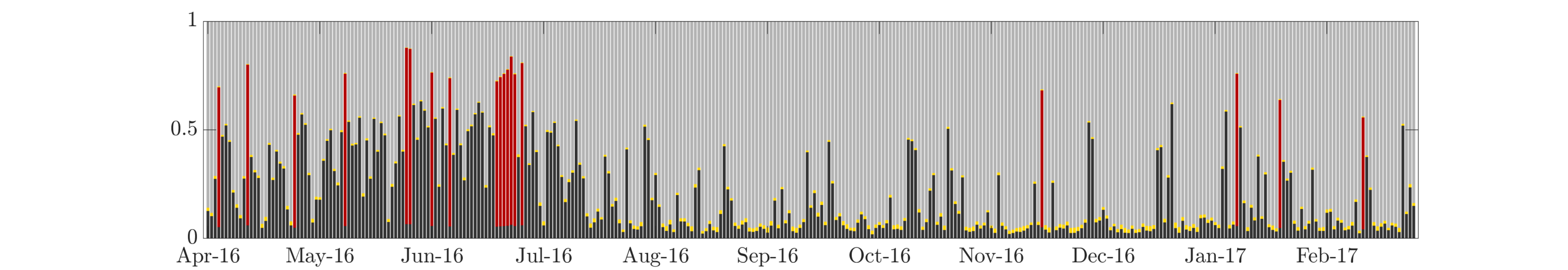}
\end{tabular}
\caption{{\protect\color{black} Daily composition of intensities in percentage: local $x_{j,t}$ (dark grey shade), amplification $x_{j,t} \xi_{j,2}$ (red), covariates $\exp(\protect\boldsymbol{\phi}_j'\mathbf{v}_{j,t})$ (gold), and  global $w_t$ (light grey).}}
\label{fig:app_risk_a0d08_intensity}
\end{figure}

The composition of total intensity for the countries under investigation is shown in Fig.~\ref{fig:app_risk_a0d08_intensity}, which reports the daily share of each component over the total.
Overall, we find that in both countries, (i) the share of local intensity, $x_{j,t}$, evolves quite smoothly over time, and (ii) social amplification, when occurring, accounts for about 70\% of the total intensity in both countries.
These results provide evidence of time variation in the drivers of country-risk perception, highlighting the role of social amplification in periods of extraordinary political events (e.g., referendums, elections, new presidents  taking office), and the relative importance of the local and global components of risk perception on ordinary days.
In particular, since August 2016 the weight of country-risk concerns, $x_{j,t}$ (dark grey), is significantly higher for the US than the UK, whereas the latter series is found to be  mainly driven by the global term, $w_t$ (light grey).
This may be a consequence of UK-related concerns experiencing a drop a few weeks after the Brexit referendum, aligning with the level of non-country-related risks. This drop may also be due to a switch in the nature of Brexit-related risks after the referendum, which started being perceived as global risks.
Overall, these results suggest that for the UK, local factors are the most influential driver of public concern only in the first part of the sample, whereas the forces driving the intensity of US concerns are more evenly distributed, with the local factor playing the major role.

To deepen the analysis of social amplification phenomena, Table~\ref{tab:app_words_regime_jump} reports the three most frequent words written during these periods (i.e., days $t$ such that $s_{j,t} \!=\! 2$).
Interestingly, we find that the these words change over time, providing evidence that social amplification is not related to a specific event or topic; instead, it rather captures swings in posting activities regarding all topical dimensions of country risks.

For example, the jump in the US series in May 2016 refers to concerns related to the possible spread of the Zika virus in the US.
Interestingly, the UK-risk perception jumped in January 2017 in response to Donald Trump taking office, thus showing that local events may have significant ripples in the public's concerns regarding other countries.
In summary, the results presented in Table~\ref{tab:app_words_regime_jump} support the hypothesis that on days with social amplification, people are more intensively communicating their concerns regarding major political events. Hence the public concerns regarding these country risks may be amplified through social media.

We extend our analysis to include Germany and China and obtained results similar to those for the US and UK. Furthermore, we used topic-based  classification of tweets to investigate political and economical risk dimensions. Interestingly, we find that the social amplification of the intensity of economic risks occurred close to the Brexit referendum, whereas Donald Trump taking office is associated with a political risk social amplification phenomenon.
We refer the reader to Section~\ref{S_sec:application_supplement} of the Supplement for further details.

\begin{table}[t!h]
\centering 
\hspace*{-8pt}
\footnotesize 
\fbox{%
\begin{tabular}{P{46pt} P{46pt} P{116pt} | P{46pt} P{46pt} P{116pt}}
\multicolumn{3}{c|}{\textbf{US}} & \multicolumn{3}{c}{\textbf{UK}} \\
\multicolumn{1}{c}{from} & \multicolumn{1}{c}{to} & \multicolumn{1}{c|}{top 3 words} & \multicolumn{1}{c}{from} & \multicolumn{1}{c}{to} & \multicolumn{1}{c}{top 3 words} \\ 
\midrule
01-Apr-16 & 01-Apr-16 & jobs, ban, lives                & 04-Apr-16 & 04-Apr-16 & cfos, brexit, country \\ 
21-Apr-16 & 21-Apr-16 & judges, liberal, forecast       & 12-Apr-16 & 12-Apr-16 & forecast, tata, steel \\ 
31-May-16 & 31-May-16 & zika, researchers, infection    & 25-Apr-16 & 25-Apr-16 & terror, new, high \\ 
12-Jun-16 & 13-Jun-16 & synthetic, drugs, pose          & 09-May-16 & 09-May-16 & forecast, may, pollution \\ 
27-Jul-16 & 27-Jul-16 & data, security, ups             & 26-May-16 & 27-May-16 & students, referendum, vote \\ 
15-Aug-16 & 15-Aug-16 & million, increase, children     & 02-Jun-16 & 02-Jun-16 & vote, global, growth \\ 
25-Aug-16 & 25-Aug-16 & just, blood, using              & 07-Jun-16 & 07-Jun-16 & forecast, big, brexit \\ 
18-Sep-16 & 18-Sep-16 & ups, early, heart               & 20-Jun-16 & 25-Jun-16 & brexit, forecast, exit \\ 
31-Dec-16 & 01-Jan-17 & obama, ideals, trump            & 27-Jun-16 & 27-Jun-16 & data, security, fire \\ 
05-Jan-17 & 06-Jan-17 & trump, data, lives              & 18-Nov-16 & 18-Nov-16 & bankers, warn, exodus \\ 
16-Jan-17 & 16-Jan-17 & trump, jobs, man                & 11-Jan-17 & 11-Jan-17 & trump, details, users \\ 
24-Jan-17 & 26-Jan-17 & fbi, hillary, grave             & 23-Jan-17 & 23-Jan-17 & free, people, high \\ 
28-Jan-17 & 04-Feb-17 & security, grid, utility         & 15-Feb-17 & 15-Feb-17 & legal, sector, people
\end{tabular}
}%
\caption{\label{tab:app_words_regime_jump} Dates of occurrence of social amplification (regime 2) and the three most frequently used words in posts of the corresponding days (excluding retweets and tweet duplicates).}
\end{table}

\begin{figure}[h!t]
\centering
\setlength{\abovecaptionskip}{3pt}
\includegraphics[trim= 43mm 0mm 11mm 0mm,clip,height= 4.6cm, width= 17.0cm]{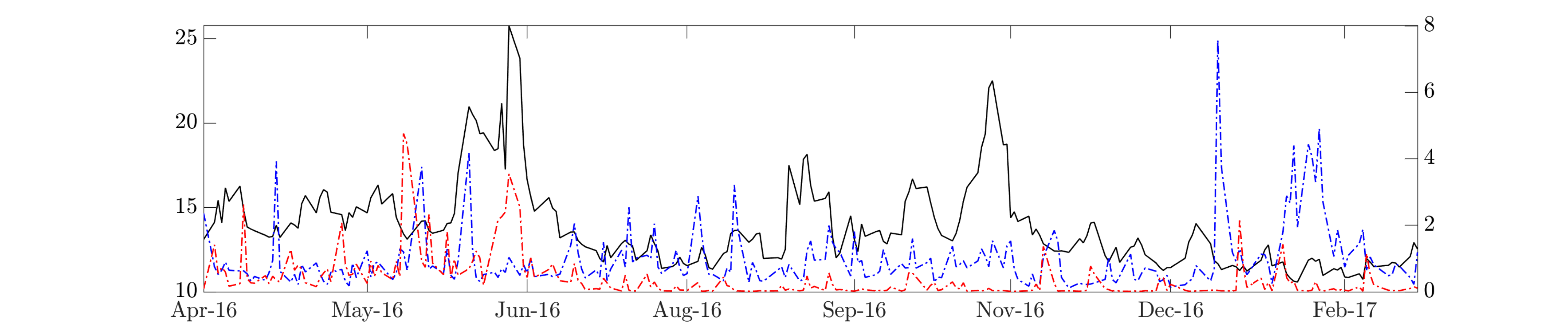} 
\caption{{\protect\color{black}VIX (solid black line, left axis) and local intensities with amplification (right axis): US (dashed blue line) and UK (dashed red line). For visualisation purposes, local intensities have been rescaled by a factor of 0.01.}}
\label{fig:app_VIX}
\end{figure}

Motivated by the long-standing debate on the relationship between risk perception and financial markets, we now investigate the performance of the extracted intensity of country-risk concerns in explaining fluctuations of a volatility index, the VIX.
This analysis offers new ways to evaluate the impact of social media amplification phenomena on market volatility.
Figure~\ref{fig:app_VIX} plots the total intensity of country-risk concerns for the US and UK (blue and red) against the VIX index (black).

To explore this relationship further, we exploit the decomposition of country risks into the persistent and social amplification components (see Eq.~\eqref{eq:latent_multiplicative_xi}) to assess the contribution of each factor in explaining fluctuations in the VIX.
Table~\ref{tab:app_local_latents_linreg_VIX_J2_Bayes} reports the estimation results of the linear regression:
\begin{equation}
\Delta VIX_t = \alpha + \boldsymbol{\beta}_{1}' \Delta \mathbf{f}_{1,t} + \ldots + \boldsymbol{\beta}_{J}' \Delta \mathbf{f}_{J,t} + \boldsymbol{\epsilon}_t, \quad \boldsymbol{\epsilon}_t \sim \mathcal{N}(0,\sigma^2),
\label{eq:linear_regression_VIX}
\end{equation}
where we assume a non-informative prior and use four alternative specifications for the vector of covariates, $\mathbf{f}_{j,t}$: (i) local intensity with amplification, $\tilde{x}_{j,t}$, (ii) jumps, $x_{j,t} \xi_{j,s_{j,t}}$, (iii) local intensity, $x_{j,t}$, (iv) local intensity and amplification, $(x_{j,t}, \, x_{j,t} \xi_{j,s_{j,t}})'$. We consider three models: two including only a single country (a-b) and another with both countries (c).
The best model for explaining fluctuations in the VIX includes the total risk intensity for both countries.
We find that including risk intensity for both countries tends to improve the model's performance, thus suggesting the existence of cross-country risk spillover effects.
Moreover, by disaggregating local intensity from the jump component, the results show that neither component in isolation has a better performance than their combination. If we also consider the results of the full model (bottom-right part of Table~\ref{tab:app_local_latents_linreg_VIX_J2_Bayes}), this may be interpreted as evidence that both components are relevant, with social amplification playing a major role.

This analysis shows that the flexible decomposition of country-risk intensity enabled by our framework allows us to investigate the impact of its components (and combinations of them) on a market index, such as the VIX.


\begin{table}[t!h]
\centering 
\scriptsize 
\fbox{%
\begin{tabular}{l *{3}{c} | l *{3}{c}} 
& (a) & (b) & (c) & & (a) & (b) & (c) \\
& US only& UK only& US, UK & & US only& UK only& US, UK\\ 
\midrule
{\scriptsize $const$} & -0.008 & -0.007 & -0.005 & {\scriptsize $const$} & -0.008 & -0.006 & -0.008 \\
& (0.081) & (0.081) & (0.080) & & (0.081) & (0.081) & (0.081) \\ 
{\scriptsize $\Delta \tilde{x}_{\scaleto{US}{3pt}}$} & \cellcolor{gray!25}0.162 &  & \cellcolor{gray!25}0.171 & {\scriptsize $\Delta x_{\scaleto{US}{3pt}}$} & 0.153 &  & 0.157 \\ 
& (0.083) &  & (0.082) &  & (0.155) &  & (0.153) \\ 
{\scriptsize $\Delta \tilde{x}_{\scaleto{UK}{3pt}}$} &  & \cellcolor{gray!25}0.232 & \cellcolor{gray!25}0.246 & {\scriptsize $\Delta x_{\scaleto{UK}{3pt}}$} &  & 0.337 & 0.324 \\ 
&  & (0.130) & (0.128) &  &  & (0.318) & (0.320) \\ 
\midrule 
{\scriptsize $DIC$} & 744.814  & 745.544  & 742.997 & {\scriptsize $DIC$}  & 747.778  & 747.635  & 748.673 \\
\midrule
\midrule
& (a) & (b) & (c) & & (a) & (b) & (c) \\
& US only& UK only& US, UK & & US only& UK only& US, UK\\ 
\midrule
{\scriptsize $const$} & -0.007 & -0.006 & -0.005 &  {\scriptsize $const$} & -0.007 & -0.005 & -0.005 \\ 
& (0.082) & (0.082) & (0.081) & & (0.082) & (0.080) & (0.081) \\ 
{\scriptsize $\Delta x_{\scaleto{US}{3pt}} \xi_{\scaleto{US}{3pt}}$} & 0.121 &  & \cellcolor{gray!25}0.136 & {\scriptsize $\Delta x_{\scaleto{US}{3pt}}$} & 0.221 &  & 0.202 \\ 
& (0.083) &  & (0.083) &  & (0.157) &  & (0.157) \\ 
{\scriptsize $\Delta x_{\scaleto{UK}{3pt}} \xi_{\scaleto{UK}{3pt}}$} &  & 0.156 & 0.180 &  {\scriptsize $\Delta x_{\scaleto{US}{3pt}}\xi_{\scaleto{US}{3pt}}$} & \cellcolor{gray!25}0.151 &  & \cellcolor{gray!25}0.159 \\ 
&  & (0.121) & (0.122) &  & (0.086) &  & (0.087) \\ 
& & & &  {\scriptsize $\Delta x_{\scaleto{UK}{3pt}}$} &  & 0.533 & 0.500 \\ 
& & & &  &  & (0.342) & (0.337) \\ 
& & & &  {\scriptsize $\Delta x_{\scaleto{UK}{3pt}}\xi_{\scaleto{UK}{3pt}}$} &  & \cellcolor{gray!25}0.226 & \cellcolor{gray!25}0.237 \\ 
& & & &  &  & (0.130) & (0.129) \\
\midrule 
{\scriptsize $DIC$}  & 746.655  & 747.019  & 746.354 & {\scriptsize $DIC$}   & 746.580  & 746.514  & 746.283
\end{tabular} 
}%
\caption{\label{tab:app_local_latents_linreg_VIX_J2_Bayes} Bayesian linear regression of VIX on \textit{local intensity with amplification}, $\Delta \tilde{x}_{j,t}$ (top-left); \textit{local intensity}, $\Delta x_{j,t}$ (top-right); \textit{jumps}, $\Delta x_{j,t} \xi_{j,s_{j,t}}$ (bottom-left); \textit{local intensity}, $\Delta x_{j,t}$, and \textit{jump}, $\Delta x_{j,t} \xi_{j,s_{j,t}}$ (bottom-right).
Standard deviations of the posterior distributions are in parentheses.
Coefficients with posterior credible intervals not containing zero are shaded in gray.}
\end{table}

\section{Conclusions} \label{sec:conclusions}

In this paper, we investigated the relation between country-risk perception and financial markets using a measure of public concern at the country level extracted from online social media data.
Count time series obtained from Web data pose several challenges, including the coexistence of jumps and smooth dynamics, with cross-sectional effects and common dynamics.
To address these challenges, we designed a novel state space framework for multivariate time series of counts that is able to account for both of these features.
The total intensity of each country series is given by the combination of a series-specific autoregressive term driving the persistent smooth dynamics, an idiosyncratic jump term that accounts for social amplification, and a global term that captures co-movements.

The proposed method is applied to count time series extracted from Twitter posts about UK and US country risks. 
Using our model, we were able to identify periods of social amplification and relate them to specific events.
Finally, to highlight the value of extracted components, we investigated their impact on the VIX and found evidence of country-risk spillover effects, driven primarily by social amplification phenomena.
%
The proposed methodology is general and has a wide range of applications in other scientific fields where count time series are popular, such as biomedical and epidemiological applications.


\section*{Acknowledgement}
Carlo Santagiustina acknowledges financial support from the European Union ODYCCEUS Horizon 2020 project, grant agreement number 732942.
{\color{black} Matteo Iacopini acknowledges financial support from the EU Horizon 2020 programme under the Marie Sklodowska-Curie scheme (grant agreement no. 887220).}

\bibliographystyle{plain}  
\bibliography{biblio_tweet.bib}

\appendix
\renewcommand{\theequation}{\thesection.\arabic{equation}}

\section{Posterior distributions} \label{sec:apdx_posterior_jumps_general}
This section reports the posterior distributions of the parameter in model \eqref{eq:model_MS_NcGa_jumps}. See the Supplement for the detailed computations.
The adaptive random walk Metropolis-Hastings (aRWMH) steps have been designed following \cite{Atchade2005adaptiveMH}, with an asymptotic acceptance rate of $\bar{r} = 0.30$.


\subsection{Sampling latents: $\mathbf{W}$, $\mathbf{X}_j$, and $\mathbf{S}_{j}$}
The trajectory of the global latent intensity $\mathbf{W}$ is estimated conditionally on $(\mathbf{Y}, \mathbf{Z}, \mathbf{V}, \boldsymbol{\phi}_z, \alpha_w, \beta_w, \delta_w, \mathbf{X}, \boldsymbol{\xi}, \mathbf{S})$. Since the filtered distribution from the state space in eq. \eqref{eq:model_MS_NcGa_jumps} is not available in closed form, we approximate it using a particle filter based on the Selection/Mutation (SM) algorithm \cite{Cappe2005HMM}.
For analogous reasons, the SISR method is used to generate the trajectory of the country-specific autoregressive latent intensity $\mathbf{X}_j$, for $j=1,\ldots,J$, conditionally on $(\mathbf{Y}_j, \mathbf{V}_j, \boldsymbol{\phi}_j, \alpha_j,\beta_j,\delta_j, \boldsymbol{\xi}_j, \mathbf{W}, \mathbf{S}_j)$.

The trajectory of the latent state variables $\mathbf{S}_{j}=(s_{j,1},\ldots,s_{j,T})'$, for each $j$, is obtained using the FFBS algorithm \cite{fruhwirth2006finite}.

\subsection{Sampling $\delta_w$ and $\delta_j$}
The posterior distribution for $\delta_w$ and $\delta_j$, for each $j=1,\ldots,J$, are
\begin{align*}
P(\delta_w | \mathbf{W}, \alpha_w, \beta_w) & \propto T\mathcal{G}a(\delta_w | a_{\delta_w},b_{\delta_w};\tau_w) \prod_{t=1}^T \textnormal{NcGa}(w_{t}|\alpha_w, \beta_w w_{t-1}, \delta_w), \\
P(\delta_j | \mathbf{X}_j, \alpha_j, \beta_j) & \propto T\mathcal{G}a(\delta_j | a_\delta,b_\delta;\tau) \prod_{t=1}^T \textnormal{NcGa}(x_{j,t}|\alpha_j, \beta_j x_{j,t-1}, \delta_j).
\end{align*}
We sample from these posterior distributions using an adaptive random walk Metropolis-Hastings (aRWMH) step with truncated lognormal proposal.

\subsection{Sampling $\alpha_w$ and $\alpha_j$}
The posterior distributions for $\alpha_w$ and $\alpha_j$, for each $j=1,\ldots,J$, are
\begin{align*}
P(\alpha_w | \mathbf{W}, \beta_w, \delta_w) & \propto \mathcal{G}a(\alpha_w | a_{\alpha_w}, b_{\alpha_w}) \prod_{t=1}^T \textnormal{NcGa}(w_{t}|\alpha_w, \beta_w w_{t-1}, \delta_w), \\
P(\alpha_j | \mathbf{X}_j, \beta_j, \delta_j) & \propto \mathcal{G}a(\alpha_j | a_\alpha,b_\alpha) \prod_{t=1}^T \textnormal{NcGa}(x_{j,t}|\alpha_j, \beta_j x_{j,t-1}, \delta_j).
\end{align*}
We sample from them using an aRWMH step with lognormal proposal.

\subsection{Sampling $\beta_w$ and $\beta_j$}
The posterior distributions for $\beta_w$ and $\beta_j$, for each $j=1,\ldots,J$, are
\begin{align*}
P(\beta_w | \mathbf{W}, \alpha_w, \delta_w) & \propto \mathcal{G}a(\beta_w | a_{\beta_w}, b_{\beta_w}) \prod_{t=1}^T \textnormal{NcGa}(w_{t}|\alpha_w, \beta_w w_{t-1}, \delta_w), \\
P(\beta_j | \mathbf{X}_j, \alpha_j, \delta_j) & \propto \mathcal{G}a(\beta_j | a_\beta,b_\beta) \prod_{t=1}^T \textnormal{NcGa}(x_{j,t}|\alpha_j, \beta_j x_{j,t-1}, \delta_j).
\end{align*}
We sample from them using an aRWMH step with lognormal proposal.

\subsection{Sampling $\eta_{j}$ and $\gamma_{j}$}
The posterior distribution for $\eta_{j}$ and $\gamma_j$, for each $j$, are given by
\begin{align*}
P(\eta_{j} | \xi_{j,2}, \gamma_{j}) & \!\propto \mathcal{G}a\Big(\! a_\eta \!+\! \gamma_{j}(c_\gamma \!+\!1), \frac{b_\eta}{1 + b_\eta \xi_{j,2}} \!\Big), \quad
P(\gamma_{j} | \xi_{j,2}, \eta_{j}) \!\propto\! \frac{\big( a_\gamma \eta_{j}^{c_\gamma +1} \xi_{j,2} \big)^{\gamma_{j}}}{\Gamma(\gamma_{j})^{b_\gamma+1}}
\end{align*}
We sample from the latter distribution via the inverse transform method.

\subsection{Sampling $\xi_{j,2}$}
We sample from the posterior distribution of the jump sizes $\xi_{j,2}$, for each $j$, using an aRWMH step with Gamma proposal
\begin{align*}
P(\xi_{j,2} & | \mathbf{Y}_{j},\mathbf{W},\mathbf{X}_{j},\eta_{j},\gamma_{j},\mathbf{S}_{j}) \!\propto\! \xi_{j,2}^{\gamma_{j}-1} \!\exp\Big( \!-\! \xi_{j,2} \Big( \! \eta_{j} \!+\!\!\! \sum_{t \in \mathcal{T}_{j,2}} \!\! x_{j,t} \Big) \Big) \!\!\prod_{t \in \mathcal{T}_{j,2}} \! (\bar{w}_t \!+\! x_{j,t}(1 \!+\! \xi_{j,2}))^{y_{j,t}}
\end{align*}

\subsection{Sampling \texorpdfstring{$\boldsymbol{\phi}_z$}{\textphi} and \texorpdfstring{$\boldsymbol{\phi}_j$}{\textphi}}
The posterior distribution of the coefficients of the covariates, $\boldsymbol{\phi}_z$ and $\boldsymbol{\phi}_j$, for $j=1,\ldots,J$, are given by
\begin{align*}
P(\boldsymbol{\phi}_z | \mathbf{Z},\mathbf{V}_z,\!\mathbf{W}) & \!\propto\! \exp\Big( \!\!-\!\frac{1}{2} (\boldsymbol{\phi}_z' \underline{\Sigma}_z^{-1} \boldsymbol{\phi}_z \!-\! 2\underline{\boldsymbol{\phi}}_z' \underline{\Sigma}_z^{-1} \boldsymbol{\phi}_z) \Big) \\
 & \cdot  \prod_{t=1}^T (w_t \!+\! \exp(\mathbf{v}_{z,t}'\boldsymbol{\phi}_z))^{z_t} \exp\big( \!-\! w_t \!-\! \exp(\mathbf{v}_{z,t}' \boldsymbol{\phi}_z) \big), \\
P(\boldsymbol{\phi}_j | \mathbf{Y}_j,\mathbf{V}_j,\mathbf{X}_j,\mathbf{W},\boldsymbol{\xi}_j,\mathbf{S}_j) &\!\propto\! \exp\Big( \!\!-\!\frac{1}{2} (\boldsymbol{\phi}_j' \underline{\Sigma}_\phi^{-1} \boldsymbol{\phi}_j \!-\! 2\underline{\boldsymbol{\phi}}' \underline{\Sigma}_\phi^{-1} \boldsymbol{\phi}_j) \Big) \\
 & \cdot \prod_{t=1}^T (\tilde{x}_{j,t} + \exp(\mathbf{v}_{j,t}'\boldsymbol{\phi}_j))^{y_{j,t}} \exp( \!- \tilde{x}_{j,t} \!-\! \exp(\mathbf{v}_{j,t}' \boldsymbol{\phi}_j) \big).
\end{align*}
To improve the mixing of the chain, we use an aRWMH step with a Normal proposal to sample element-wise from the posterior distribution of $\boldsymbol{\phi}_z$ and $\boldsymbol{\phi}_j$.

\subsection{Posterior for $\Lambda_{j}$}
The posterior distribution for each row $l$ of each transition matrix $\Lambda_j$ is 
\begin{align*}
P(\boldsymbol{\lambda}_{j,l}|\mathbf{S}_j) & \propto \mathcal{D}ir(\underline{\lambda}_{j1} + N_{l1}(\mathbf{S}_j), \ldots, \underline{\lambda}_{jL} + N_{lL}(\mathbf{S}_j)).
\end{align*}

\end{document}